\pdfoutput=1 

\documentclass[reprint,aps,prl,superscriptaddress,showpacs]{revtex4-1}

\usepackage{amsmath,amssymb}
\usepackage{dcolumn}
\usepackage{multirow}
\usepackage[colorlinks=true,linkcolor=blue,urlcolor=blue,citecolor=blue]{hyperref}
\usepackage{graphicx}

\newcommand{\BINP}{Budker Institute of Nuclear Physics, 630090 Novosibirsk, Russia}
\newcommand{\ANL}{Argonne National Laboratory, Argonne, Illinois 60439, USA}
\newcommand{\NSU}{Novosibirsk State University, 630090 Novosibirsk, Russia}
\newcommand{\TPU}{Tomsk Polytechnic University, 634050 Tomsk, Russia}
\newcommand{\Nikhef}{Nikhef, 1098 XG Amsterdam, Netherlands}

\newcolumntype{d}[1]{D{.}{.}{#1}}

\DeclareMathOperator{\re}{Re} 

\begin{document}

\title{Measurement of the two-photon exchange contribution to the elastic $e^{\pm}p$~scattering cross sections at the VEPP-3 storage ring}

\author{I.\,A.~Rachek}\email{I.A.Rachek@inp.nsk.su}\affiliation{\BINP{}}
\author{J.~Arrington}\affiliation{\ANL{}}
\author{V.\,F.~Dmitriev}\affiliation{\BINP{}}\affiliation{\NSU{}}
\author{V.\,V.~Gauzshtein}\affiliation{\TPU{}}
\author{R.\,E.~Gerasimov}\affiliation{\BINP{}}\affiliation{\NSU{}}
\author{A.\,V.~Gramolin}\email{A.V.Gramolin@inp.nsk.su}\affiliation{\BINP{}}
\author{R.\,J.~Holt}\affiliation{\ANL{}}
\author{V.\,V.~Kaminskiy}\affiliation{\BINP{}}
\author{B.\,A.~Lazarenko}\affiliation{\BINP{}}
\author{S.\,I.~Mishnev}\affiliation{\BINP{}}
\author{N.\,Y\lowercase{u}.~Muchnoi}\affiliation{\BINP{}}\affiliation{\NSU{}}
\author{V.\,V.~Neufeld}\affiliation{\BINP{}}
\author{D.\,M.~Nikolenko}\email{D.M.Nikolenko@inp.nsk.su}\affiliation{\BINP{}}
\author{R.\,S\lowercase{h}.~Sadykov}\affiliation{\BINP{}}
\author{Y\lowercase{u}.\,V.~Shestakov}\affiliation{\BINP{}}
\author{V.\,N.~Stibunov}\affiliation{\TPU{}}
\author{D.\,K.~Toporkov}\affiliation{\BINP{}}\affiliation{\NSU{}}
\author{H.~\lowercase{de}~Vries}\affiliation{\Nikhef{}}
\author{S.\,A.~Zevakov}\affiliation{\BINP{}}
\author{V.\,N.~Zhilich}\affiliation{\BINP{}}\affiliation{\NSU{}}

\begin{abstract}
The ratio of the elastic $e^+ p$ to $e^- p$ scattering cross sections has been measured precisely, allowing the determination of the two-photon exchange contribution to these processes. This neglected contribution is believed to be the cause of the discrepancy between the Rosenbluth and polarization transfer methods of measuring the proton electromagnetic form factors. The experiment was performed at the \mbox{VEPP-3} storage ring at beam energies of 1.6 and 1.0~GeV and at lepton scattering angles between $15^{\circ}$ and $105^{\circ}$. The data obtained show evidence of a significant two-photon exchange effect. The results are compared with several theoretical predictions.
\end{abstract}


\maketitle

The proton is a fundamental building block of \mbox{matter}. In order to understand its complex internal structure and the interaction between its constituents, quarks and gluons, it is crucial to have reliable knowledge of the proton electromagnetic form factors~\cite{ARNPS.54.217, JPhysG.34.S23, PPNP.59.694, FewBodySyst.46.1, PRL.111.101803}.

In the spacelike region, these form factors are measured using elastic electron-proton scattering. For a long time, the only experimental method available was the Rosenbluth method based on the following well-known formula describing the unpolarized elastic $ep$~scattering cross section in the one-photon exchange approximation:
\begin{equation}
\frac{d \sigma}{d \Omega_e} = \frac{1}{\varepsilon (1 + \tau)} \bigl[\varepsilon G_E^2 (Q^2) + \tau G_M^2 (Q^2)\bigr] \frac{d \sigma_{\text{Mott}}}{d \Omega_e}, \label{eq1}
\end{equation}
where $\varepsilon = \left[1 + 2(1 + \tau) \tan^2{(\theta_e / 2)}\right]^{-1}$ is the virtual photon polarization parameter, $\theta_e$~is the electron scattering angle, $\tau = Q^2 / (4M^2)$, $Q^2$~is the four-momentum transfer squared, $M$~is the proton mass, $G_E (Q^2)$ and $G_M (Q^2)$ are the proton electric and magnetic form factors, and $d \sigma_{\text{Mott}} / d \Omega_e$ is the Mott differential cross section. 

Another method of measuring the ratio $G_E / G_M$, the so-called polarization transfer method, was proposed back in 1968~\cite{SovPhysDokl.13.572, SovJPartNucl.4.277}, but implemented only several decades later. Unexpectedly, a clear discrepancy was observed at $Q^2 \gtrsim 1~\text{GeV}^2$ between the results obtained by these two methods~\cite{PRL.84.1398, PRC.64.038202, PRL.88.092301, PRC.68.034325, PRC.71.055202, PRL.104.242301, PRC.85.045203}. This contradictory situation has attracted great attention since it raises questions about the entire methodology of electron scattering experiments.

It has been suggested that the origin of the discrepancy is the failure of the one-photon exchange approximation to properly describe the results of unpolarized measurements, and that the two-photon exchange (TPE) effect should be taken into account~\cite{PRL.91.142303, PRL.91.142304, ARNPS.5.171, PPNP.66.782}. The leading TPE contribution is due to interference between the one-photon and two-photon exchange amplitudes, $\mathcal{M}_{1\gamma}$ and $\mathcal{M}_{2\gamma}$. The latter is usually represented as a sum of ``soft'' (calculated in the soft-photon approximation) and ``hard'' parts, $\mathcal{M}_{2\gamma} = \mathcal{M}_{2\gamma}^{\text{soft}} + \mathcal{M}_{2\gamma}^{\text{hard}}$~\cite{PPNP.66.782, JPhysG.41.115001}. The soft part is infrared divergent and independent of the proton structure, while the hard part is finite and highly model dependent. The standard prescriptions~\cite{RMP.41.205, PRC.62.054320} for radiative corrections (RCs) take into account only the portion of $\mathcal{M}_{2\gamma}^{\text{soft}}$ needed to cancel the corresponding infrared divergences due to bremsstrahlung. Note that such a separation of $\mathcal{M}_{2\gamma}$ into soft and hard parts is ambiguous. In this Letter, we follow the Mo--Tsai convention~\cite{RMP.41.205}.

There are many attempts to calculate $\mathcal{M}_{2\gamma}^{\text{hard}}$, but as the results are model dependent and often conflicting, experimental data are required. Since the interference TPE term changes sign depending on the charge sign of the scattered particle, the TPE effect can be studied by comparison under similar experimental conditions of elastic electron-proton and positron-proton scattering. Such measurements were performed in the 1960s~\cite{PR.128.1842, PR.139.B1079, PRL.17.407, PLB.25.242, PRL.19.1191, PR.166.1336, PLB.26.178, PRL.21.482}, but their precision is insufficient to reach any definitive conclusion~\cite{PRC.69.032201}. To fill this gap, there are two other new experiments~\cite{PRC.88.025210, NIMA.741.1} in addition to the reported measurement at the \mbox{VEPP-3} storage ring (Novosibirsk, \mbox{Russia}).

The experimentally measured quantity is the ratio $R = \sigma (e^+p) / \sigma (e^- p)$ of the elastic $e^+ p$ and $e^- p$ scattering cross sections. The desired hard TPE contribution to Eq.~(\ref{eq1}),
\begin{equation}
\delta_{2\gamma} = \frac{2\re{\bigl(\mathcal{M}_{1\gamma}^{\dagger} \mathcal{M}_{2\gamma}^{\text{hard}}\bigr)}}{|\mathcal{M}_{1\gamma}|^2}, \label{eq2}
\end{equation}
can be determined from~$R$ after taking into account the first-order RCs~\cite{JPhysG.41.115001}. Finally, the results are presented as the ratio $R_{2\gamma} = (1 - \delta_{2\gamma}) / (1 + \delta_{2\gamma})$.

The experiment had two data-taking runs: run~I at a beam energy of about 1.6~GeV and run~II at 1.0~GeV. The average beam current was about 20~mA. Electron and positron beams were alternated regularly during the data collection, so that each experimental cycle with both beam polarities took approximately 1~hour. We performed about 3000 such cycles during the entire experiment and collected integrated luminosities of 320 and $600~\text{pb}^{-1}$ in run~I and run~II, respectively.

The experiment used an internal gas target, based on an open-ended storage cell with an elliptical cross section of $13 \times 24~\text{mm}^2$ and a length of 400~mm. High-purity hydrogen gas was injected into the cell center to provide a target thickness of~$\approx 10^{15}~\text{atom}/\text{cm}^2$. Four cryopumps served to remove the hydrogen gas flowing from the cell ends to the vacuum chamber. The pressure in the center of the storage cell during target operation was about \mbox{$1.5 \times 10^{-3}$}~Torr.

\begin{figure}[b]
\includegraphics[width=\linewidth]{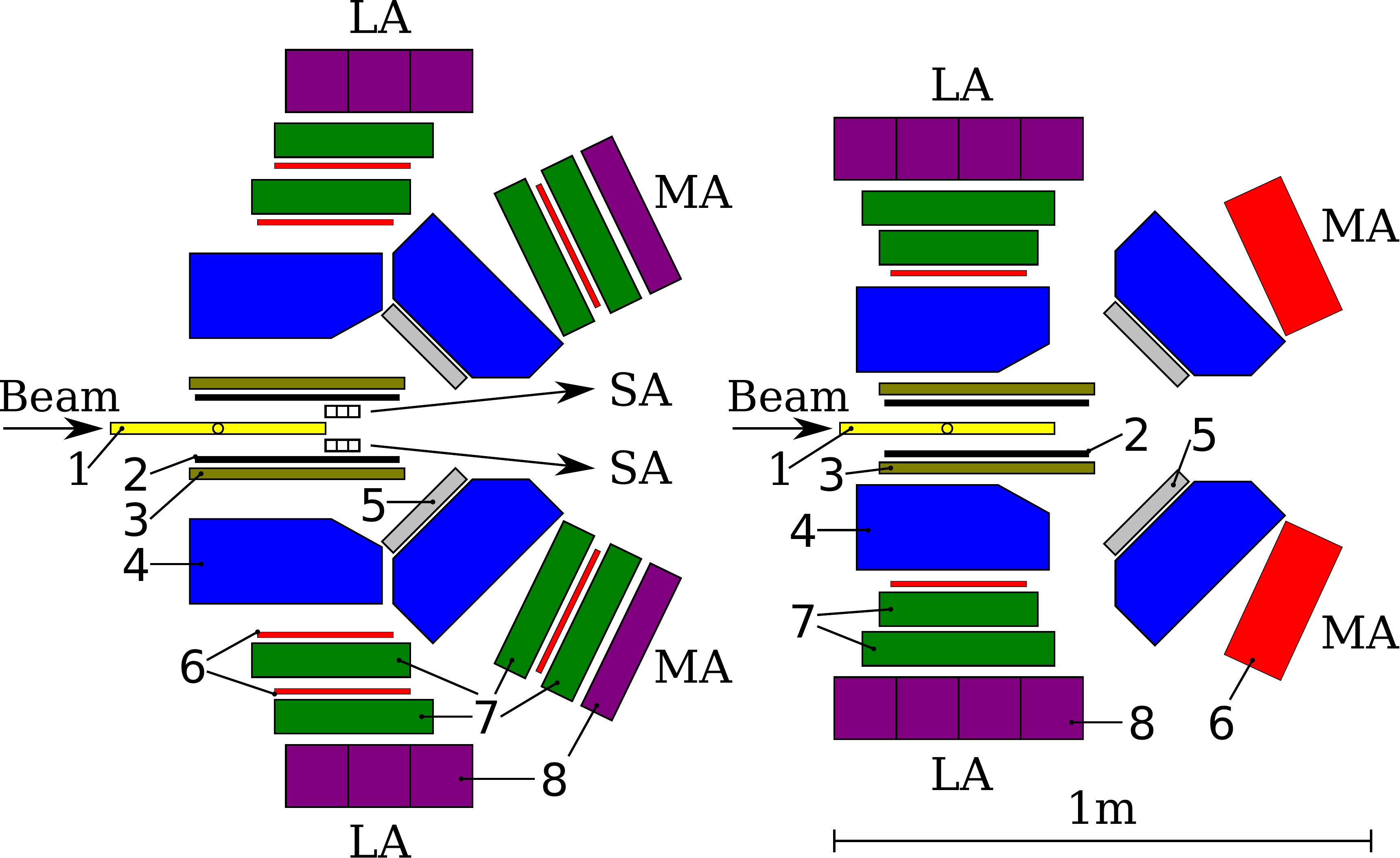}
\caption{\label{fig1}(color online) The detector configurations for run~I and run~II (left and right panels, respectively). Labels: 1---storage cell; 2---beryllium sheet; 3---multiwire proportional chamber; 4---drift chamber; 5---acrylic glass; 6---plastic scintillator; 7---CsI crystals; 8---NaI crystals; SA, MA, LA---detector arms.}
\end{figure}

The scattered lepton (electron or positron) and the recoil proton were detected in coincidence by a wide-aperture nonmagnetic detector (see Fig.~\ref{fig1}). This was composed of two nearly identical sectors, upper and lower, placed symmetrically with respect to the median plane of the storage ring. The azimuthal acceptance of each sector was $60^{\circ}$.

As can be seen from Fig.~\ref{fig1}, the detector had two different configurations in run~I and run~II. In the first run, there were three ranges of the lepton scattering angle: \mbox{$7^{\circ}$--$16^{\circ}$} (small angles, SAs), $15^{\circ}$--$28^{\circ}$ (medium angles, MAs), and $55^{\circ}$--$83^{\circ}$ (large angles, LAs), corresponding to three pairs of detector arms. The SA arms were used to detect scattered leptons only, while the MA and LA arms detected both leptons and protons.

The LA and MA arms included two multiwire proportional chambers and four drift chambers for charged particle tracking, four segmented electromagnetic calorimeters comprised of CsI and NaI crystals for lepton energy measurements, and six plastic scintillators for event triggering and proton identification. The radiation lengths of the calorimeters were about $10.6 X_0$ for each LA arm and $8.3 X_0$ for each MA arm. Two multilayer tungsten-scintillator sandwich calorimeters each with a radiation length of~$8.6 X_0$ were used in the SA arms.

In run~II, there were only two scattering angle ranges used: $15^{\circ}$--$30^{\circ}$ (MA) and $65^{\circ}$--$105^{\circ}$ (LA). The LA arms were positioned at more backward angles. The radiation lengths of the LA calorimeters were the same as in run~I. The MA arms were equipped with two thick plastic scintillators installed in place of the crystal calorimeters.

Additionally, 6~mm thick beryllium sheets and 30~mm thick acrylic glass (see Fig.~\ref{fig1}) were placed in front of the wire chambers to shield them from the large background of low-energy electrons.

The SA events of run~I and MA events of run~II were only used for luminosity normalization. It is commonly believed that for the corresponding forward-angle kinematics ($Q^2 \approx 0.1~\text{GeV}^2$ and $\varepsilon > 0.9$) the hard TPE effect is small~\cite{PPNP.66.782, JPhysG.40.115003}. We can assume therefore that $R_{2\gamma}$ is very close to unity in this case.

To select elastic scattering events, the following kinematic correlations were used: between the polar angles of the lepton and proton; between their azimuthal angles; between the polar angle and energy of the lepton and proton; and between the lepton scattering angle and the proton energy. Different combinations of the corresponding kinematic cuts were applied to the LA, MA, and SA events. Additionally, time-of-flight measurements and $dE/dx$ analysis were used for proton identification.

A detailed \textsc{Geant4} simulation was performed to take into account RCs and to estimate the background from pion-production reactions. The processes $e p \rightarrow e' n \, \pi^+$, $e p \rightarrow e' p \, \pi^0$, $\gamma^* p \rightarrow n \pi^+$, and $\gamma^* p \rightarrow p \, \pi^+ \pi^-$ were simulated using an event generator based on the MAID2007 and \mbox{2-PION}-MAID models~\cite{MAID}. According to the simulation, the fraction of the background events among the selected ones does not exceed 4\% for the LA ranges of both runs and is negligible for the MA and SA ranges.

To account for the first-order RCs, the \mbox{ESEPP} event generator~\cite{JPhysG.41.115001, ESEPP} was used. The following options of \mbox{ESEPP} were chosen: the dipole parametrization for the proton form factors; an accurate QED calculation beyond the soft-photon approximation for first-order bremsstrahlung; the vacuum polarization correction that includes the hadronic contribution; and the soft TPE terms according to Mo and Tsai~\cite{RMP.41.205}.

Note that the interference term between lepton and proton bremsstrahlung has opposite signs in the cases of $e^- p$ and $e^+ p$ scattering, and thus affects the measured ratio~$R$. This effect is comparable in size with the hard TPE effect under study. Unfortunately, proton bremsstrahlung cannot be calculated in a model-independent way. We used the model~\cite{JPhysG.41.115001}, which goes beyond the usual soft-photon approximation, but still assumes that the intermediate hadronic states are the virtual-proton ones.

Because of bremsstrahlung, RCs strongly depend on the specific kinematic cuts used to select events. The angular cuts that we applied can be characterized by the inequalities $\left||\phi_e - \phi_p| - \pi\right| < \Delta \phi$ and $|\theta_p - \theta_p^*| < \Delta \theta$. Here, $\phi_e$ and $\phi_p$ are the azimuthal angles of the lepton and proton, $\theta_p$ is the polar angle of the proton, and $\theta_p^*$ is the expected value of~$\theta_p$, calculated from $\theta_e$ and $E_{\text{beam}}$ assuming elastic kinematics. Another kinematic cut affecting RCs is the cut on the scattered lepton energy. This can be expressed in the form $E_{\theta} - E_{\text{cal}} < \Delta E$, where $E_{\theta}$ is calculated from~$\theta_e$ and $E_{\text{cal}}$ is determined from the energy deposition in the calorimeter. In our case, the energy cut is conservative and RCs are determined mainly by the angular cuts.

Several factors allowed us to reduce the systematic uncertainties of the measurement. In particular, the nonmagnetic detector ensured identical acceptances for electrons and positrons. Its symmetric configuration helped to suppress the negative effects due to possible displacement and slope of the beam with respect to the \mbox{VEPP-3} median plane. Additionally, the target thickness and the integrated beam current were eliminated from consideration by the luminosity normalization. Finally, the frequent alternation of the beam polarities suppressed errors due to slow variations in time of the detection efficiency.

The systematic errors coming from different sources are listed in Table~\ref{tab1} for the four kinematic points at which the ratio $R_{2\gamma}$ is reported. The points No.1 and No.2 correspond, respectively, to the LA and MA events of run~I, and the points No.3 and No.4 represent the LA events of run~II divided into two bins.

\begingroup
\squeezetable
\begin{table}[b]
\caption{\label{tab1}Contributions to the systematic error of $R_{2\gamma}$ (\%).}
\begin{ruledtabular}
\begin{tabular}{ld{3}d{3}d{3}d{3}}
& \multicolumn{2}{c}{Run~I} & \multicolumn{2}{c}{Run~II} \\
\cline{2-3}\cline{4-5} \\[-2mm]
& \multicolumn{1}{c}{No.1} & \multicolumn{1}{c}{No.2} & \multicolumn{1}{c}{No.3} & \multicolumn{1}{c}{No.4} \\
\hline \\[-2mm]
(1) Unequal beam energies & 0.024 & 0.015 & 0.014 & 0.014 \\
(2) Unequal beam positions & 0.162 & 0.172 & 0.047 & 0.017 \\
(3) Unequal detection efficiencies & 0.055 & 0.055 & 0.031 & 0.031 \\
(4) Kinematic cuts & 0.207 & 0.019 & 0.022 & 0.022 \\
(5) Background subtraction & 0.140 & 0.050 & 0.070 & 0.050 \\
(6) Radiative corrections & 0.090 & 0.050 & 0.130 & 0.040 \\
Total systematic error, $\Delta R_{2\gamma}^{\text{syst}}$ & 0.32 & 0.20 & 0.16 & 0.08 \\
\end{tabular}
\end{ruledtabular}
\end{table}
\endgroup

The first three contributions shown in Table~\ref{tab1} are because of slightly different experimental conditions during the data collection with electron and positron beams. The contribution~(1) is very small due to accurate real-time measurements of the beam energy using a Compton backscattering setup~\cite{JINST.9.T06006}. The beam position was determined by three different methods: using tracking data from the detector to reconstruct the event vertex; using data from the \mbox{VEPP-3} pickup electrodes; and using movable beam scrapers to probe the beam position. As a result, the relative positions of electron and positron beams were known with an accuracy of 0.07 and 0.15~mm for the horizontal and vertical directions, respectively. These uncertainties give the second contribution. The third one is mainly due to variations in time of the tracking efficiency, not suppressed completely by alternating the beam polarities. To estimate this effect, we studied the fraction of events with reconstructed tracks among the coincidence events, selected without using any information from the wire chambers.

The contributions (4), (5), and (6) are because of imperfections in the event selection and data analysis. The first of them was estimated by varying the kinematic cuts and then subtracting the corresponding contribution of statistical fluctuations. Another one arises from the background subtraction procedure. Finally, the uncertainty in RCs is due to their model dependence on the form factor parametrization used and the neglect of higher-order bremsstrahlung~\cite{JPhysG.41.115001}.

All of these factors, listed in Table~\ref{tab1}, affect the luminosity normalization. In fact, the errors at the luminosity normalization points (LNPs) and at the points No.1--No.4 caused by the effects (1), (2), and (3) are correlated. For this reason, we included all LNP errors in the systematic errors given in Table~\ref{tab1}. Similarly, statistical uncertainties due to luminosity normalization are incorporated into the statistical uncertainties of~$R_{2\gamma}$ at the points No.1--No.4.

\begingroup
\squeezetable
\begin{table}[b]
\caption{\label{tab2}Parameters and results of the experiment.}
\begin{ruledtabular}
\begin{tabular}{ld{1.4}d{1.4}d{1.3}d{1.4}d{1.4}d{1.3}}
& \multicolumn{3}{c}{Run~I} & \multicolumn{3}{c}{Run~II} \\
\cline{2-4} \cline{5-7} \\[-2mm]
& \multicolumn{1}{c}{No.1} & \multicolumn{1}{c}{No.2} & \multicolumn{1}{c}{LNP} & \multicolumn{1}{c}{No.3} & \multicolumn{1}{c}{No.4} & \multicolumn{1}{c}{LNP} \\
\hline \\[-2mm]
$E_{\text{beam}}$ (GeV) & 1.594 & 1.594 & 1.594 & 0.998 & 0.998 & 0.998 \\
$\varepsilon_{\text{min}}$ & 0.29 & 0.89 & 0.96 & 0.18 & 0.33 & 0.88 \\
$\varepsilon_{\text{max}}$ & 0.58 & 0.97 & 0.99 & 0.33 & 0.51 & 0.97 \\
$\langle \varepsilon \rangle$ & 0.452 & 0.932 & 0.980 & 0.272 & 0.404 & 0.931 \\
$\langle Q^2 \rangle$ $(\text{GeV}^2)$ & 1.51 & 0.298 & 0.097 & 0.976 & 0.830 & 0.128 \\
$\langle \theta_{e} \rangle$ & \multicolumn{1}{l}{$66.2^{\circ}$} & \multicolumn{1}{l}{$20.8^{\circ}$} & \multicolumn{1}{l}{$11.4^{\circ}$} & \multicolumn{1}{l}{$91.3^{\circ}$} & \multicolumn{1}{l}{$75.4^{\circ}$} & \multicolumn{1}{l}{$21.4^{\circ}$} \\
$\Delta \phi$, $\Delta \theta$ & \multicolumn{1}{l}{$3.0^{\circ}$} & \multicolumn{1}{l}{$5.0^{\circ}$} & \multicolumn{1}{c}{$\cdots$} & \multicolumn{1}{l}{$3.0^{\circ}$} & \multicolumn{1}{l}{$3.0^{\circ}$} & \multicolumn{1}{c}{$\cdots$} \\
$\Delta E / E_{\theta}$ & 0.25 & 0.45 & \multicolumn{1}{c}{$\cdots$} & 0.29 & 0.29 & \multicolumn{1}{c}{$\cdots$} \\
$N_{\text{sim}}^+ / N_{\text{sim}}^0$ & 1.0347 & 1.0600 & \multicolumn{1}{c}{$\cdots$} & 1.0501 & 1.0206 & \multicolumn{1}{c}{$\cdots$} \\
$N_{\text{sim}}^- / N_{\text{sim}}^0$ & 0.9981 & 1.0563 & \multicolumn{1}{c}{$\cdots$} & 1.0117 & 0.9898 & \multicolumn{1}{c}{$\cdots$} \\
$R$ & 1.0705 & 1.0037 & \multicolumn{1}{c}{$\cdots$} & 1.0555 & 1.0447 & \multicolumn{1}{c}{$\cdots$} \\
$R_{2\gamma}$ & 1.0332 & 1.0002 & \multicolumn{1}{c}{1} & 1.0174 & 1.0133 & \multicolumn{1}{c}{1} \\
$\Delta R_{2\gamma}^{\text{stat}}$ & \pm 0.0112 & \pm 0.0012 & \multicolumn{1}{c}{$\cdots$} & \pm 0.0049 & \pm 0.0037 & \multicolumn{1}{c}{$\cdots$} \\
$\Delta R_{2\gamma}^{\text{syst}}$ & \pm 0.0032 & \pm 0.0020 & \multicolumn{1}{c}{$\cdots$} & \pm 0.0016 & \pm 0.0008 & \multicolumn{1}{c}{$\cdots$} \\
\end{tabular}
\end{ruledtabular}
\end{table}
\endgroup

Table~\ref{tab2} provides the experimental results: the values of $R_{2\gamma}$ with the total statistical and systematic uncertainties. These results are obtained assuming that $R_{2\gamma}$ is equal to unity at the normalization points ($R_{2\gamma}^{\text{LNP}} = 1$). Also listed are the kinematic parameters of the measurement, the $\Delta \phi$, $\Delta \theta$, and $\Delta E$ cuts, the raw ratios~$R$, and the quantities $N_{\text{sim}}^{+} / N_{\text{sim}}^{0}$ and $N_{\text{sim}}^{-} / N_{\text{sim}}^{0}$ obtained in the \textsc{Geant4} simulation and needed to extract $R_{2\gamma}$~\cite{JPhysG.41.115001}.

\begin{figure*}
\centering
\includegraphics[width=0.45\textwidth,clip]{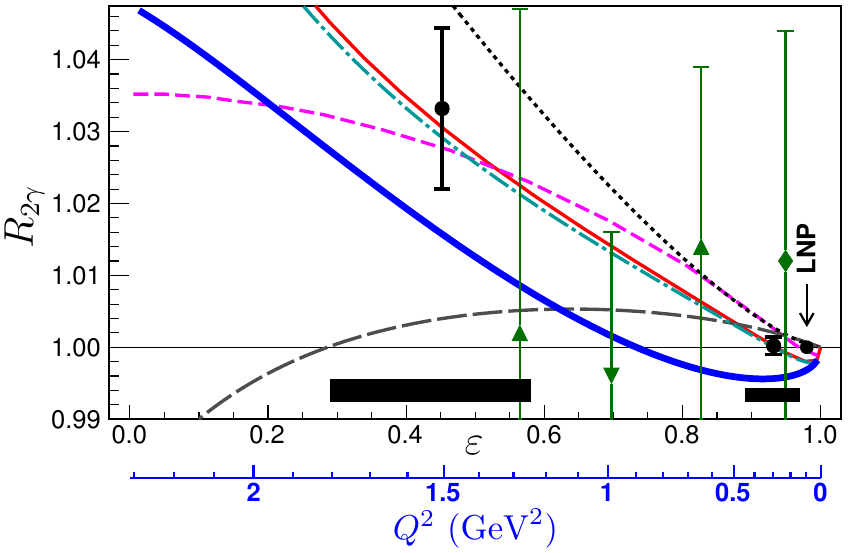}
\hspace{5mm}
\includegraphics[width=0.45\textwidth,clip]{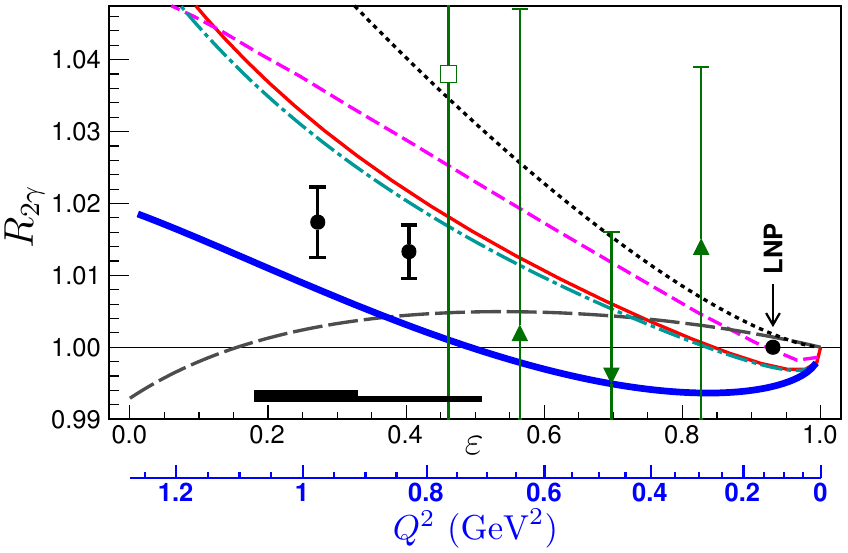}
\caption{\label{fig2}(color online) Experimental data (points) and some predictions (curves) for the ratio $R_{2\gamma}$ as a function of~$\varepsilon$ or~$Q^2$. The left and right panels correspond, respectively, to run~I and run~II. Data points:
open square \cite{PR.139.B1079},
closed inverted triangle \cite{PRL.17.407},
closed diamond \cite{PLB.25.242},
closed triangle \cite{PR.166.1336}, and
closed circle---this experiment.
Error bars of our points (closed circles) are related to the statistical uncertainties; the shaded bands show the total systematic uncertainty and the bin size for each data point. The curves are from 
Ref.~\cite{PRC.78.025208} (cyan dash-dotted line),
\cite{PRC.72.034612} (red thin solid line),
\cite{PRC.90.015206} (blue thick solid line),
\cite{PAN.76.937} (gray long-dashed line),
\cite{PRC.70.028203} (magenta short-dashed line), and
\cite{PRC.84.054317} (black dotted line).}
\end{figure*}

Figure~\ref{fig2} compares our results with some of the existing experimental data~\cite{PR.139.B1079, PRL.17.407, PLB.25.242, PR.166.1336} and several theoretical or phenomenological predictions~\cite{PRC.78.025208, PRC.72.034612, PRC.90.015206, PAN.76.937, PRC.70.028203, PRC.84.054317}. Only those of the old data points which approximately correspond to our kinematics, defined in Fig.~\ref{fig2} by the beam energy and $\varepsilon$~values, are shown. It can be seen that our results are in agreement with the previous measurements, but significantly more precise. The figure also shows that the hadronic calculations, Refs.~\cite{PRC.78.025208, PRC.72.034612}, are in good agreement with the data of run~I, but overestimate the values of~$R_{2\gamma}$ obtained in run~II. In contrast, the phenomenological fit~\cite{PRC.90.015206} underestimates $R_{2\gamma}$ at all the measured points. Note that this fit has been corrected by us to switch from the Maximon--Tjon prescription~\cite{PRC.62.054320} for the soft TPE terms, used in Ref.~\cite{PRC.90.015206}, to the Mo--Tsai prescription~\cite{RMP.41.205}, used by us (see Ref.~\cite{JPhysG.41.115001} for details). It should be emphasized that the models~\cite{PRC.78.025208, PRC.72.034612, PRC.90.015206} resolve the form factor discrepancy at high $Q^2$ values by taking into account the hard TPE effect. The other three predictions~\cite{PAN.76.937, PRC.70.028203, PRC.84.054317} are worse in overall agreement with our data.

\begingroup
\squeezetable
\begin{table}[b]
\caption{\label{tab3}Comparison of our results with predictions.}
\begin{ruledtabular}
\begin{tabular}{lccccc}
& \multirow{2}{*}{$R_{2\gamma}^{\text{LNP}}$} & \multirow{2}{*}{\normalsize $\bigl(\frac{\chi^2}{n_{\text{d.f.}}}\bigr)$} & \multicolumn{2}{c}{$R_{2\gamma}^{\text{LNP}}$} & \multirow{2}{*}{\normalsize $\bigl(\frac{\chi^2}{n_{\text{d.f.}}}\bigr)$} \\
\cline{4-5} \\[-2mm]
& & & \multicolumn{1}{c}{Run~I} & \multicolumn{1}{c}{Run~II} & \\
\hline \\[-2mm]
Borisyuk and Kobushkin~\cite{PRC.78.025208} & 1 & 2.14 & 0.9979 & 0.9972 & 3.80 \\
Blunden {\it et~al}.~\cite{PRC.72.034612} & 1 & 2.94 & 0.9980 & 0.9974 & 4.75 \\
Bernauer {\it et~al}.~\cite{PRC.90.015206} & 1 & 4.19 & 0.9969 & 0.9946 & 1.00 \\
Tomasi-Gustafsson {\it et~al}.~\cite{PAN.76.937} & 1 & 5.09 & 1.0007 & 1.0014 & 5.97 \\
Arrington and Sick~\cite{PRC.70.028203} & 1 & 7.72 & 0.9995 & 0.9996 & 8.18 \\
Qattan {\it et~al}.~\cite{PRC.84.054317} & 1 & 25.0 & 1.0005 & 1.0018 & 22.0 \\
No hard TPE ($R_{2\gamma} \equiv 1$) & 1 & 7.97 & 1 & 1 & 7.97 \\
\end{tabular}
\end{ruledtabular}
\end{table}
\endgroup

Our results can also be renormalized according to the tested model. In this case, the values of~$R_{2\gamma}$ at the points No.1--No.4 should be multiplied by the corresponding values of $R_{2\gamma}^{\text{LNP}}$ predicted by the model. This is illustrated in Table~\ref{tab3}, where the normalization coefficients for each of the predictions~\cite{PRC.78.025208, PRC.72.034612, PRC.90.015206, PAN.76.937, PRC.70.028203, PRC.84.054317} are given. Also shown are the chi-square values per degree of freedom, $\chi^2 / n_{\text{d.f.}}$, characterizing the agreement between the prediction and the data. The second and the third columns correspond to the normalization to unity, while the next three columns correspond to the normalization in accordance with the predictions. The last row of Table~\ref{tab3} refers to the case of the hard TPE contribution being zero. It can be seen that this case is not consistent with our data. Note also that the fit~\cite{PRC.90.015206} has a large change in the chi-square value with the change in normalization, showing a very good agreement in the case of normalization to the predicted values of~$R_{2\gamma}^{\text{LNP}}$.

The conclusion that the predictions~\cite{PRC.78.025208, PRC.72.034612, PRC.90.015206} seem the most plausible remains valid regardless of the normalization used. Nevertheless, an accurate normalization of our data is desired and can be achieved later if new precise measurements or reliable calculations of the hard TPE effect at $Q^2 \approx 0.1~\text{GeV}^2$ become available.

In summary, the first high-precision measurement of the hard TPE contribution to the elastic $e^{\pm} p$~scattering cross sections has been performed. The results obtained show evidence of a significant hard TPE effect. They are in moderate agreement with several TPE predictions explaining the form factor discrepancy at high $Q^2$ values. Therefore, our data support the suggestion that the discrepancy is due to the neglected hard TPE contribution to elastic electron-proton scattering.

\begin{acknowledgments}
The authors are grateful to V.\,S.~Fadin, \mbox{A.\,L.~Feldman}, and A.\,I.~Milstein for fruitful discussions. \mbox{P.\,G.~Blunden} and E.~Tomasi-Gustafsson are thanked for providing us with their calculations. We acknowledge the \mbox{VEPP-3} staff for the stable operation of the storage ring during the experiment. This work was supported by the Grant Council of the President of the Russian Federation (Grant No.~MK-525.2013.2), the Ministry of Education and Science of the Russian Federation (Contract No.~02.740.11.0245.1), the Russian Foundation for Basic Research (Grants No.~12-02-00560, No.~12-02-33140, No.~13-02-00991, and No.~13-02-01023), and the US Department of Energy (Grant No.~DE-AC02-06CH11357).
\end{acknowledgments}

\end{document}